\documentclass{IEEEtran}
\usepackage[utf8]{inputenc}
\usepackage{graphicx}
\usepackage{varwidth}
\usepackage{xcolor}
\usepackage{url}

\title{Data-driven Software Security:\\ Models and Methods}
\author{}

\author{\IEEEauthorblockN{\'Ulfar Erlingsson}\\
\IEEEauthorblockA{Google, Inc.}}

\begin{document}

\maketitle

\begin{abstract}
For computer software, our security models, policies, mechanisms, and means of assurance were primarily conceived and developed before the end of the 1970's. However, since that time, software has changed radically: it is thousands of times larger, comprises countless libraries, layers, and services, and is used for more purposes, in far more complex ways. It is worthwhile to revisit our core computer security concepts. For example, it is unclear whether the Principle of Least Privilege can help dictate security policy, when software is too complex for either its developers or its users to explain its intended behavior.

One possibility is to take an empirical, data-driven approach to modern software, and determine its exact, concrete behavior via comprehensive, online monitoring. Such an approach can be a practical, effective basis for security---as demonstrated by its success in spam and abuse fighting---but its use to constrain software behavior raises many questions. In particular, three questions seem critical. First, can we efficiently monitor the details of how software \emph{is} behaving, in the large?  Second, is it possible learn those details without intruding on users' privacy?  Third, are those details a good foundation for security policies that constrain how software \emph{should} behave?  
    
This paper outlines what a data-driven model for software security could look like, and describes how the above three questions can be answered affirmatively. Specifically, this paper briefly describes methods for efficient, detailed software monitoring, as well as methods for learning detailed software statistics while providing differential privacy for its users, and, finally, how machine learning methods can help discover users' expectations for intended software behavior, and thereby help set security policy. Those methods can be adopted in practice, even at very large scales, and demonstrate that data-driven software security models can provide real-world benefits.
\end{abstract}

\section{Introduction}

The core concepts of computer security have remained mostly the same since they were developed and articulated in the 1970’s and late 1960’s, e.g., in the seminal papers on protection by Lampson and by Saltzer and Schroeder~\cite{Protection,SaltzerAndSchroeder}.
However, since then, surprisingly little progress has been made in increasing the practical security of computer users, especially compared to the exponential improvements in other aspects of computing during the same time.
Indeed, by most measures, the problems of computer security have grown progressively worse over those decades, even though a plethora of computer security mechanisms has been developed and deployed, and the use of technologies such as advanced cryptography has become ubiquitous.

Today’s computer users suffer every year some inconvenience or damage as a result of computer security issues, as their personal information is stolen from their devices or vulnerable databases, as their computers are infected with unwanted software or harvested into botnets, or simply as their credit cards or vehicles are recalled because of a security breach or vulnerability.
It’s therefore worthwhile to consider approaches to computer security based on new foundations---although any such effort must acknowledge that security is too intricate and pervasive a problem to admit a panacea.

This short position paper presents a \emph{data-driven software security model} that is founded on an abstraction of \emph{empirical programs}, which pairs software with data on all security-relevant events in all of its execution traces.

Enforcing this data-driven model will, for example, 
naturally
disallow network access by the Microsoft Windows Solitaire game~\cite{Solitaire}
and also disallow messages that trigger the Heartbleed vulnerability~\cite{Heartbleed}.
In the model, security policies are automatically derived from historical evidence,
and the Solitaire game doesn't use the Windows networking libraries, even though it includes them,
and no naturally-occurring TLS heartbeat messages with huge payloads have ever existed.

This  paper  further motivates  and describes this data-driven model of software security, and gives examples of practical, useful methods for its application.

\subsection{Security Models and the Difficulty of Setting Policies}

It is not a new observation that we have made disappointingly little progress in secure computing.  
In a series of talks and papers around the turn of the millennium, Butler Lampson, one of the founding figures in computer security, gave a good overview of the field, its aspirations and promise, and its failures in practice~\cite{LampsonRealWorld}.
In this work, 
Lampson convincingly explains the difficulties of computer security, e.g., how it is even harder than traditional security because universal networking has created an unbounded set of would-be attackers, and because computers’ precise, faithful execution implies that exploits can only be stopped by fully guaranteeing the absence of vulnerabilities.

\begin{figure}[t]
\centering
\begin{tabular}{lcl}
Security policy & $=$ & Functional specification \\
Security mechanism & $=$ & Software implementation \\
Security assurance & $=$ & Program correctness \\
\emph{Security model} & $=$ & \emph{Programming methodology}
\end{tabular}
\caption{The correspondence between aspects of computer security and computer software, the last one an addition to those identified by Lampson~\cite{LampsonRealWorld}.}\label{fig:terminology}
\end{figure}

\begin{figure*}[t]
\centering
\fbox{\begin{varwidth}{\dimexpr\textwidth-2\fboxsep-2\fboxrule\relax}
{\em Permit only low-level executions that programmers intended to be possible, \\
     unless given explicit, special permission.}
\end{varwidth}}
\caption{The general maxim of the widely-used programmer-intent software security model, as applied to security-relevant events during software execution.}\label{fig:piss}
\end{figure*}

Most importantly, Lampson draws the correspondence, shown in Figure~\ref{fig:terminology}, between security policies, mechanisms, and assurance and the more general pursuit of program correctness (i.e., establishing that software correctly implements a specification).
To his list, Figure~\ref{fig:terminology} adds a correspondence between security models, such as access-control lists, capabilities, or information-flow tracking, and different programming methodologies, such as imperative or functional programming, logical or declarative formalisms, or reactive software modules.
Like programming paradigms, 
security models can be best seen as alternative means towards the same end
(at least, if we follow Lampson and ignore general non-interference, etc.~\cite{hyperprop}).
Tasks and approaches easy in one Turing-complete language
may be difficult in another,
and the same holds true for security models,
which, in general,
can restrict executions only to the same set
of enforceable security policies~\cite{Enforceable}.

Lampson's correspondence highlights how the definition of the right security policy---which can be enforced with assurance by the selected mechanisms---is the key obstacle to creating secure software.
Software developers are famously loath to fully specify their intended functionality, and infamously likely to get such specifications wrong, when forced to create them; furthermore, only recently has it become practically feasible to guarantee program correctness for simple software~\cite{SeL4}.
Figure~\ref{fig:terminology} identifies security policy as just a form of specification,
albeit one that is particularly crucial and hard to get right.
It can only be in vain hope that we ask software developers, administrators, or users to define security policy and select enforcement mechanisms, 
when the more fundamental, and much better understood, task of creating software that correctly performs its task has proven to be such an insurmountable challenge.

Depending on the task at hand, and the guarantees needed, certain security models and mechanisms may be particularly well-suited to provide assurance, just as software may be best implemented using a certain methodology.
Security policies can range widely in their intent and granularity, 
like any other form of software specification,
and for some purposes it can be simple to derive and set the intended policy.
For example, 
simple prohibitions of permitted information-flow can comprehensively protect secrecy,
just as declaratively implementing functionality in Datalog can fully ensure decidability.
Security models that enable simple, useful security policies for large classes of software are clearly advantageous, even though their simplicity will prevent such models from addressing many real-world security concerns.

A particularly important 
security model that admits simple policies is one that encompasses mechanisms to thwart the exploits of low-level software vulnerabilities.

\subsection{Automatically-derived Policies for Low-level Software}
Over the last couple of decades, stack-based buffer overflows and memory-corruption vulnerabilities have become a primary exploit vector and a critical software security issue.

In defending against such attacks, the security model outlined in Figure~\ref{fig:piss} has been particularly effective at defining simple, useful security policies that successfully prevent exploits.
In this security model---here termed ``programmer-intent software security''---security policies are automatically derived from software source code (or binaries) by identifying simple program properties that are obviously true, based on the programming-language abstractions and semantics and the clear intent of the programmers.

This security model has been instantiated multiple times, for example by placing canaries or cookies on the execution stack to preserve the programmer-intended integrity of function return-address values, or by introducing artificial heterogeneity and randomness to capture an intent that programs be insensitive to the concrete representation of values like pointers~\cite{lowLevelSoftwareSecurity}.
Those instantiations 
have adhered to the maxim of Figure~\ref{fig:piss}
and have had to explicitly permit only a handful of special 
cases, e.g., for dynamic loading and libraries and signal delivery.
Unfortunately, many of those instantiations tie policy and mechanism too closely and intricately together for the underlying model to be clearly identifiable.

The clearest examples of this security model is the work to enforce the programmer's intended control and data flow that is often termed Control-Flow Integrity (CFI) and Data-Flow Integrity (DFI)~\cite{programShepherd,CFI,SafeCODE,DFI,WIT,fwdCFI}.
In that work, it is clear how different security policies of varying granularity can be automatically derived from the software itself, and how those policies can (greatly) constrain the attacker from exploiting low-level vulnerabilities.
Those policies can
operate at different levels of abstraction and
aim to enforce a detailed abstract model of the program with full precision,
or make only binary, coarse-grained distinctions between code or data,
or even leave certain activity completely unconstrained.
For example, some CFI mechanisms apply only to C++ VTables, but do so precisely, whereas others apply to all indirect control flow in a very coarse manner;
similarly, data-flow integrity mechanisms differ in their abstractions, with some based on data allocation but others on read/write patterns, etc.

\begin{figure*}[t]
\centering
\fbox{\begin{varwidth}{\dimexpr\textwidth-2\fboxsep-2\fboxrule\relax}
{\em Permit only executions that historical evidence shows to be common enough,\\
     unless given explicit, special permission.}
\end{varwidth}}
\caption{The general maxim of the data-driven software security model of this paper, as applied to security-relevant events during software execution.}\label{fig:ddss}
\end{figure*}

The great advantage of these control-flow and data-flow security mechanisms
is that their policy is dictated by the already-written software program.
In no case
is the user required to specify policy details: they must simply choose between security policies with different enforcement mechanisms, and different levels of assurance.
Of course, that choice may still be challenging, because a wide variety of mechanisms exists---ranging from binary-translation emulators to highly-optimized compiler-inserted inline guards---with greatly differing software-engineering and performance characteristics.
Furthermore, the parameters of each mechanism are likely tweakable, e.g., to change the level of precision or make other tradeoffs.
Finally, only some of those mechanisms will attempt to provide high assurance, such as those that verify the static CFI properties of the final binary~\cite{Strato}.
However, these are easily-made, commonplace engineering tradeoffs, compared to the primary obstacle of writing security policies or program specifications identified by Lampson.

Mechanisms that embody this programmer-intent security model are now used near universally, in one form or another, undoubtedly due to their lack of user-specified security policies.
However, at best, application of this model removes just one set of vulnerabilities---approximately the ones eliminated by a rigorously type-safe programming language---and leaves unremedied other vulnerabilities, such as actual logic errors made by programmers.
It is worthwhile to consider other ways in which useful, practical security policies can be defined without user specification, e.g., by automatically basing such policies on the experience gathered from absolutely all executions of a software program.

\section{A Data-driven Software Security Model}
Let's posit that software programs could be accompanied by a comprehensive summary of \emph{all} executions of that program, detailing the program's historical behavior in every instance.
Indeed, let's define a new abstraction of an \emph{empirical program} as the static software program (e.g., its source code or executable text) coupled with the multiset of \emph{all} of its execution traces.
Those traces would be captured at some level of granularity
and,
for the purposes of security,
may include only security-relevant execution events.

Since modern software is generally online---to receive security updates, if nothing else---the practical implementation of this empirical program abstraction is not a farfetched idea.

\begin{figure}[t]
\centering
\includegraphics[width=\columnwidth]{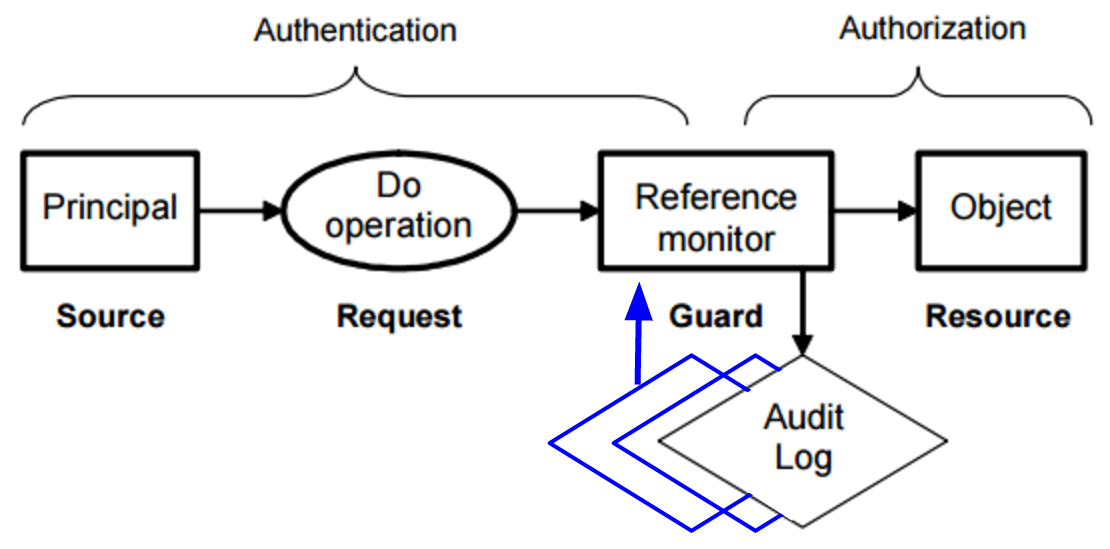}\vspace*{-1.4em}
\caption{Data-driven software security tied to access control, as per~\cite{LampsonRealWorld}.}\label{fig:ddsacls}
\end{figure}

Such an abstraction would naturally support the security model outlined in Figure~\ref{fig:ddss}, which simply prohibits all novel, security-relevant behavior, unless especially permitted.
This model could, by default, prevent many software attacks,
such as privilege-escalation exploits 
of the vulnerabilities
regularly discovered in esoteric operating system services.
Most recently,
this model's enforcement would have blocked exploits of the CVE-2016-0728 vulnerability
by prohibiting use of the Linux \texttt{keyctl} system call\label{ref:keyctl}
in commonly-used applications,
since historical evidence 
would have shown that this software never used \texttt{keyctl} or kernel keyrings. 

\begin{figure}[t]
\vspace*{0.43in}
\centering
\includegraphics[width=\columnwidth]{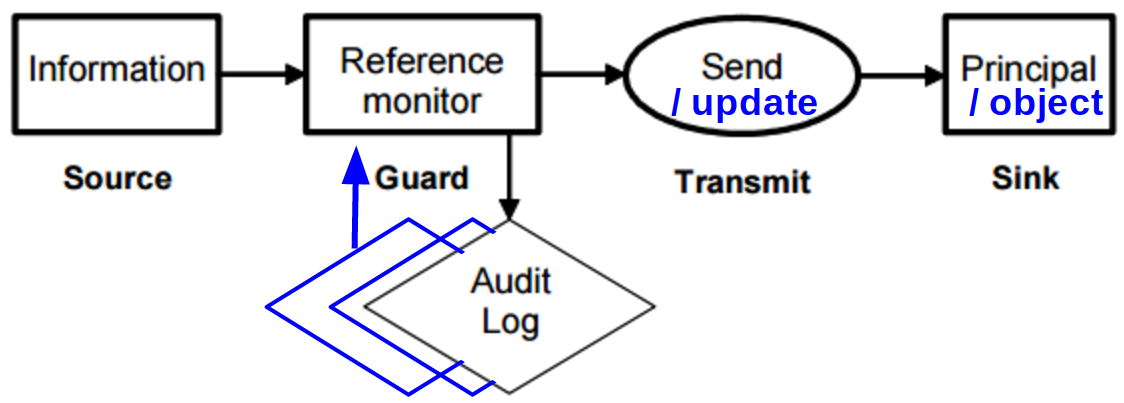}\vspace*{-1em}
\caption{Data-driven software security tied to information flow, as per~\cite{LampsonRealWorld}.}\label{fig:ddsif}
\end{figure}

Of course, to apply this model in a practical security enforcement mechanism, there are a number of obstacles and naturally-arising questions.
For example, what security policy should be enforced on the first program execution?
%
%
But, whatever the obstacles,
the data-driven security model of Figure~\ref{fig:ddss}
is attractive for many reasons.
In particular, 
it seems like a natural basis for software security enforcement
to consider how that software has behaved in the past, overall
(cf., state-based security enforcement
that considers only the current execution~\cite{hbac}).
Also,
as shown in Figures~\ref{fig:ddsacls} and~\ref{fig:ddsif},
this model can be naturally combined with existing security models,
by simply ensuring that operations proceed and information flows
in accordance with historical audit logs 
(thereby raising the importance of audit to parity with 
the other aspects of Lampson's Gold Standard~\cite{LampsonRealWorld}).

Crucially, this data-driven security shares with programmer-intent security
the great advantage 
that policy is primarily dictated by the already-written (and -executed) empirical programs.
Therefore,
policies can be simple and easy to specify.

The primary parameters of data-driven security policies would be the event
abstraction used in execution traces---for example,
network service requests, API or system calls, function calls, or simply security privileges---as
well as
the historical frequency by which 
security-relevant events must have been seen,
to be permitted in the current execution.
Although historical evidence may be interpreted in myriad ways
(e.g., using elaborate machine learning),
efficiency and robustness concerns are likely to favor simpler methods
(e.g., utilizing the set of observed events, not their sequences).

For clarity,
this position paper considers
only security-related events, such as system calls,
which can be easily summarized and combined using set-theoretic operations.
An event is considered to be supported by 
historical evidence
if it has occurred at least once (or $k$ times for some fixed, low threshold $k$)
in the execution traces;
otherwise, it is prohibited by the security policy.
Security policies
that such sets of permitted operations to reduce software attack surface 
have proven to be of great practical value,
e.g., in network firewalls and operating-system sandboxing~\cite{ChromeSandbox}.

%
Simple, data-driven software security policies
can be particularly well suited
to large-scale, popular software applications,
which are typically
bloated in size,
composed of innumerable platforms, modules, and libraries,
and full of arcane or unused functionality.
In such commonly-used software
there exist many dusty corners with lurking vulnerabilities,
as well as embedded interpreters, dynamic-library loaders, and reflection APIs
that attackers can exploit to perform arbitrary behavior.
Simply disallowing previously-unseen security-relevant events
can effectively thwart such attacks,
and for widely-used software
that can be a strong basis of historical evidence 
from an abundance of execution traces.
%
%
%

Of course, data-driven security is not a panacea, 
and its limitations, obstacles, and open questions must be carefully considered.
However, 
as outlined above,
data-driven security enforcement can provide significant benefits,
at least for the commonly-used software that most affects users.

\subsection{Differences from Anomaly and Intrusion Detection}
The most prominent past attempts 
at automatically deriving security policies from execution traces
have centered on the
techniques of anomaly detection and software intrusion detection.
Although the details differ,
the common idea in this work is
to use a set of traces collected from benign training runs (or trial runs)
to determine what constitutes ``normal'' executions,
and enforce compliance to this security policy 
during a later operational phase~\cite{Denning,AnomalySurveyTwo}.
Because 
software tends to have a long tail of possible behaviors
and
these techniques use only a partial set of training traces
(e.g., from a special training phase),
they invariably suffer from a great number of false reports
when deployed in practice.
As a result,
they have seen limited practical use,
except as voluminous but helpful warning notifications for manual operators.

\paragraph*{Comprehensiveness}
This paper's data-driven software security model 
relies on the empirical program abstraction to
avoid such falsely-reported security violations.
By definition, empirical programs include \emph{all} execution traces,
not just those from training runs.
Those traces should also include all executions 
performed during the software's development and testing,
which should ensure that any latent, actual software feature will be represented,
even for the first use of unpopular software.
The long-tail behavior of commonly-used software
should be particularly transparent---or so it would seem---since
their policies could summarize
billions of execution traces.

\paragraph*{Level of abstraction}
For the above to hold true,
the abstractions and methods used for data-driven security policies  
must be chosen carefully.
They cannot, in particular, be so fine-grained
that they include user-specific behavior that is orthogonal to the software's semantics.
For example,
security policies 
based on the content of user input to text-editing software,
or the URL bar of a Web browser,
would be likely guaranteed to trigger false errors in the future.
Therefore,
data-driven security is best based on simpler, coarser abstractions,
such as sets of system calls used in this paper,
perhaps augmented with invariants on parameter values and their magnitudes.
Techniques for selecting such abstractions 
can be
built on the foundation of inferring convergent software invariants~\cite{Daikon,Diduce}.

\paragraph*{Development-process integration}
Also,
data-driven security techniques must be 
at least partially integrated into engineering processes
and preferably used throughout the software development lifecycle.
For example,
test coverage analysis could be used to determine
when 
security policies are comprehensive and permissive enough to 
allow wide software deployment.
Fuzzing, concolic execution, or other automated testing techniques 
could be used to increase coverage (e.g., as in~\cite{MiningSandboxes}).
However, 
care would have to be taken, or software 
with embedded interpreters or other 
general-purpose modules
might be found to exhibit all possible behaviors.

Such software-engineering integration
would also be invaluable in maintaining security policies
as software is updated for security, stability, or behavior 
(which can be frequent in modern software),
or is given the occasional major upgrade.
(Of course, 
a proper accounting of software updates
would require redefining 
the empirical program abstraction
to account for differences in versions over time.)
Most importantly,
the maintenance of data-driven security policies would have to be
integrated into software development,
with the deployment of new, summarized security policy
treated as a form of software security update.

\paragraph*{Real-world execution traces}
No matter how software development is performed,
there is no substitute for 
the evidence from real-world execution traces
in the crafting of data-driven security policies.
Testing is at best partial, 
and software is often used for unanticipated purposes
in unexpected configurations;
no testing framework can hope to 
replicate all end-user activity and the environments
that may affect software execution.
It is inevitable that
some execution paths
will be seen first during real-world execution
(e.g., it's not hard to imagine that 
this might be true for
software-emulated denormal floating-point computations).
Furthermore,
it may be important 
that security policy disallow some behavior
that is common during software development,
such as debugger- or automation-driven inputs;
otherwise,
data-driven security policies
might, by default, open the door to
cases like the well-known sendmail Debug script vulnerability, CVE-1999-0095.

\paragraph*{Context-specificity}
To increase the precision of security policies,
execution traces can
include information about environmental factors,
such as time, locale, user preferences, etc.,
which may affect software behavior in myriad ways.
For example,
software execution may be highly dependent on time:
the Y2K bug and its effect comes to mind,
as does end-of-year processing in some business software.
In this case,
one could hope this behavior had been
captured in execution traces during functionality or integration testing.
More problematically, some software may contain functionality ``easter eggs''
that were hidden even during software development.
In this latter case,
one can ask whether software behavior that has never been seen before, even during testing,
should not simply be counted as a bug, resulting in program termination;
certainly, software crashes often---without good reason---and here at least there would be a good reason.

\paragraph*{Remediation options}
More generally,
a range of remedial actions may be taken 
upon violation of a data-driven security policy.
Fail-stop enforcement that halts execution
may be practical in many cases,
e.g., for non-critical software that can be restarted by the user,
and where data loss is not a concern,
as long as the software can be reset to avoid an infinite loop of halting.
Also,
it may sometimes be practical to simply ask the user whether to proceed,
and summarize their decisions to set future policy;
a variant of this approach is used for security-relevant permissions on some mobile phone and Web platforms.
Finally,
even if execution is silently allowed to proceed,
both system administrators and software developers
could make good use of high-accuracy reports of software exploits,
e.g., to set firewall rules or develop security updates.
Remediation by silent alerts would blur the distinction
with traditional anomaly detection,
except that data-driven security enforcement
should trigger vastly fewer alerts.

\paragraph*{Deployment bootstrapping}
Even when using silent alerts,
there should be no enforcement of
data-driven security policies
until
those policies have converged and stabilized,
through the summarization of sufficient real-world execution traces.
This is a key 
distinction between
the data-driven software security model
and traditional anomaly detection.
By integrating with the software-development lifecycle,
including initial real-world use of the software,
the process of converging to stable security policies 
can be monitored
and at some point---after careful consideration using domain knowledge---the
switch must be flipped to start enforcing those policies.

Thus, 
data-driven enforcement won't likely bring any security benefits
during software development, testing, and early deployments;
its benefits will accrue primarily after software has become frequently-used enough,
which, fortunately, is also when its improved security will impact the most users.

\subsection{Open Questions and Formal Modeling}
There still remain open questions
about this data-driven software security model that cannot be answered here.

For example, if security policy is driven by real-world execution traces,
it is not difficult to imagine that attackers might try to get malicious behavior
classified as benign, e.g., by using a Sybil attack to facilitate Mimicry attacks~\cite{Sybil,Mimicry}.
In other contexts, such as crowd-sourced restaurant reviews,
those attacks are  largely prevented by risk analysis of registered user accounts.
However, those attacks might be devastating in a software context,
especially where there is no well-defined, accountable registry of users.
Even so,
in some domains, such as datacenter computing,
this obstacle can clearly be overcome,
e.g., by eliminating Sybils or managing their number.

There also remain more formal questions 
about the model,
especially as regards to its empirical program abstraction.
From a formal-language perspective,
programs can be seen as language recognizers,
and from this viewpoint
the insecurity of
modern software partially stems from it recognizing too large a set of inputs.
An empirical program restricts the set of recognized inputs 
by disallowing some events in execution traces,
at some levels of abstraction,
while the static text of the software itself
implicitly defines a subset of allowed events. 
Clearly,
it would be helpful in applying the data-driven software security model
if there was a sound basis for formal reasoning about empirical programs.

However,
it is first necessary to establish
that comprehensive software execution tracing can be done, efficiently enough, in practice,
and used to derive data-driven policies that provide useful security benefits.

\section{Methods for Data-driven Software Security}
Despite its relative simplicity and other attractive qualities,
a data-driven software security model 
is not likely to be straightforward to apply, in most domains.
For example,
both
the crucial integration with software-engineering processes
and the
deployment of mechanisms to construct and maintain security policy
are likely to be major challenges, by themselves.
Also, 
it is still more of an art than science to
select the abstractions, granularity, and thresholds of empirical programs,
and to determine when trace data has converged into useful security policies---requiring 
the skills of hard-to-find artisan security engineers
that are domain experts.

Even so,
over the last few years,
at Google
we have constructed
several data-driven security mechanisms, and have utilized them in various ways,
as part of software products and production infrastructure.
Many of those mechanisms have been experimental, but some have seen significant deployment.

This section describes three of those mechanisms,
which establish that
execution traces can be collected
with low-enough overhead and in a way that protects end-user privacy,
and that those traces can be used to protect users' security and privacy in novel ways.

\subsection{Efficient Monitoring of Software Execution Details}
At Google, 
in work led by Michael Vrable,
we have considered 
data-driven security policies about system-call behavior
for production software that runs in our datacenters.
For such software, the empirical program abstraction can 
be relatively easily realized
by integrating with Google's test-driven development
and by collecting execution-trace summaries 
from thousands of process instances.
Also,
fail-stop enforcement can be particularly 
well-suited to fault-tolerant software,
which is designed to gracefully tolerate process failures
and automatically discard requests that trigger such failures.

Our experience has shown that system-call-trace-based security policies
can be efficiently collected, summarized, and enforced
using standard technologies,
such as \texttt{ptrace} and \texttt{seccomp\_bpf} on Linux---neither
of which incurs any significant per-system-call cost,
even if applied holistically as might by done with a system-wide profiler~\cite{DCPI}.

We have developed further techniques for efficient tracing 
at the level of functions, library routines, or network messages, 
based on reordering executable-binary code (including message-marshaling code)
such that execution of supposedly-unused code
can be blocked
using operating-system memory protection.
Finally, we have created mechanisms to robustly handle
abrupt, unexpected changes in software behavior,
and the resulting storms of events and other disruptions.
Therefore,
the data-driven security model can be realized efficiently, in practice,
at least for policies that involve sets of system calls or sets of other support routines and services.

\begin{figure}[t]
\centering
\includegraphics[width=\columnwidth]{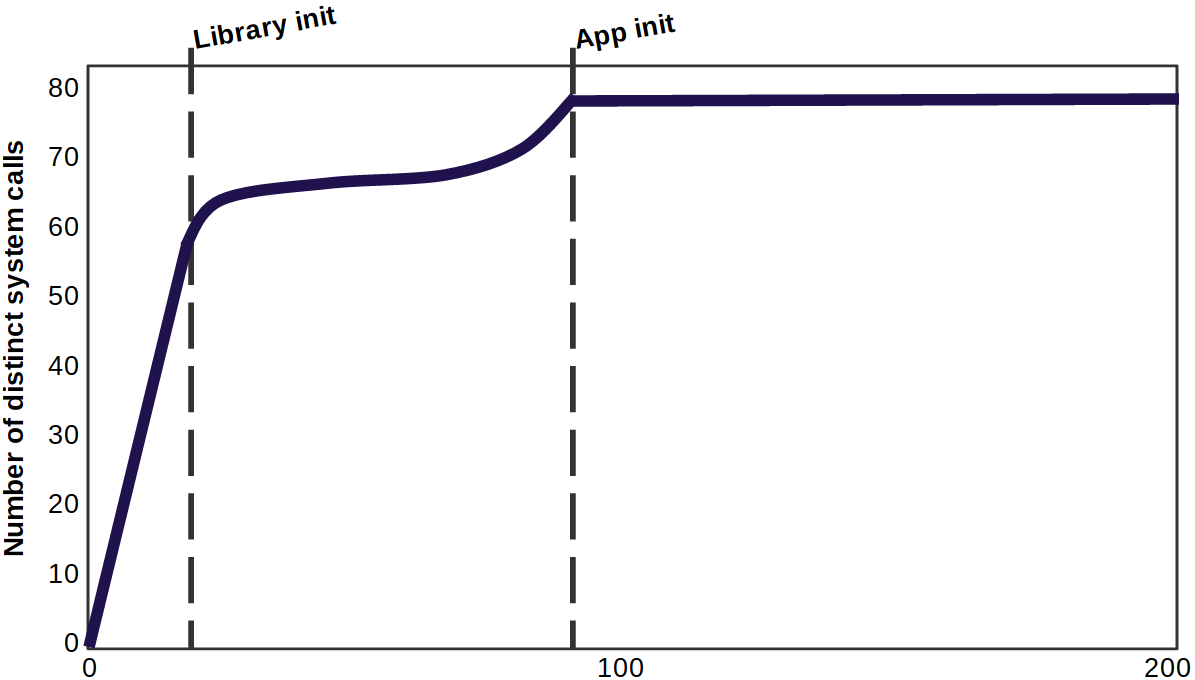}
\includegraphics[width=\columnwidth,trim=4 0 0 0]{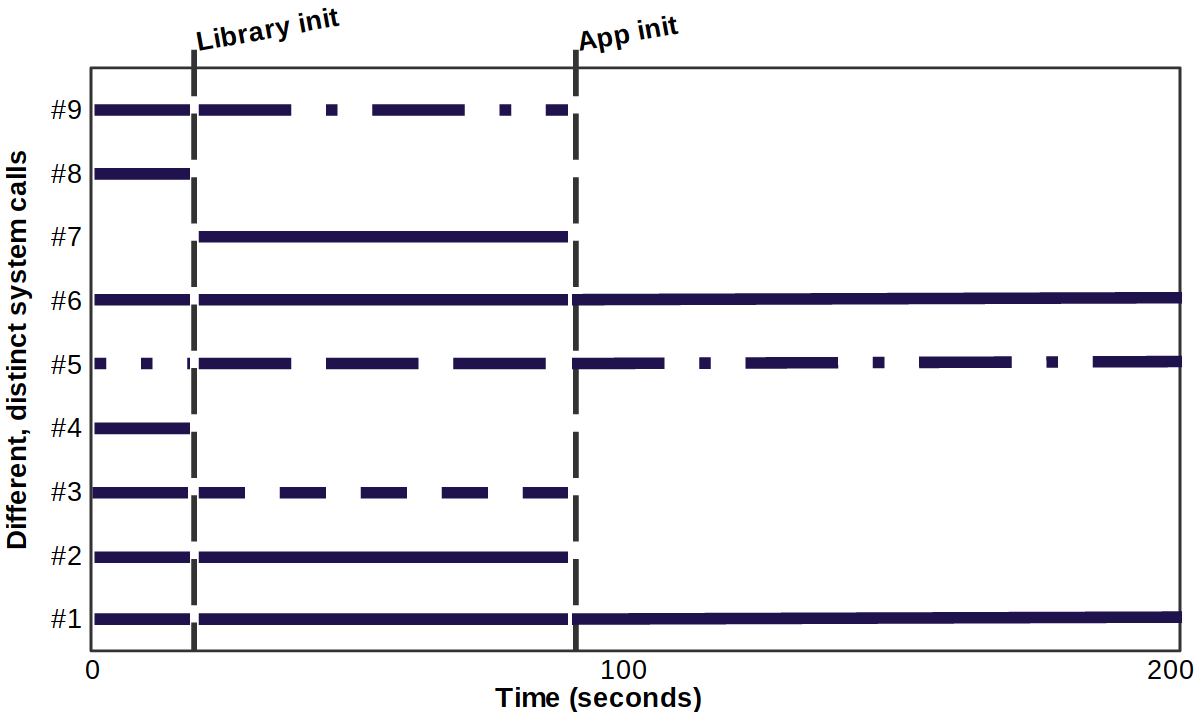}
\caption{
A representation of the system calls in an empirical program of a large, feature-rich software service,
drawn from concrete data on the first 200 seconds after startup. 
Above, the cumulative number of system calls used converges to fewer than 80, out of the 300+ in Linux.
Below, 
system call use varies between initialization phases, 
with very few calls used in the final production-service phase
(only the first 9 system calls are represented).}\label{fig:chips}
\end{figure}

Furthermore, our work has established
that enforcement of such policies can significantly reduce the attack surface
that is exposed to potentially-vulnerable software.
When viewed as an empirical program,
even the most feature-rich software
makes a relatively modest use of underlying system services---although
the software may include code, modules, and libraries for
all possible functionality, as well as several kitchen sinks.

For example, 
the upper half of
Figure~\ref{fig:chips}
shows how, empirically, one such program uses less than a third of the Linux system calls.
The lower half of Figure~\ref{fig:chips} also shows
how such software can exhibit phase-specific behavior
that can be used to improve security,
e.g., by further constraining policy after initialization,
much as is done by the Chrome Web browser~\cite{ChromeSandbox}.
Notably,
in this case,
the software made no use of the \texttt{keyctl} call,
so an automatically-derived security policy
would have prevented exploits of the CVE-2016-0728 vulnerability discussed on page~\pageref{ref:keyctl}.
\subsection{Privacy-preserving Learning of Software Execution Data}
Software monitoring can have privacy consequences for users,
even if it is not immediately apparent how the data is privacy sensitive.
Simply knowing that a software feature has been active can have significant consequences---as
is effectively demonstrated
in television courtroom dramas,
where perpetrators are frequently found guilty 
based on evidence from log records captured for benign purposes.

There are many possible methods 
that can protect the privacy and anonymity of software users
participating in the collection of execution traces for empirical programs.
For example, 
data can be collected and combined using elaborate cryptographic methods---such
such as partially-homomorphic Paillier encryption---or users can simply 
rely on Tor-like network anonymization
when providing trace data.

At Google,
we have developed and deployed
some particularly attractive 
privacy-preserving technologies for monitoring client data
based on
the the ideas of randomized response~\cite{rappor}.
These technologies are available in the open-source RAPPOR project,
to be found at \texttt{\url{https://github.com/google/rappor}}~\cite{rappor}.
RAPPOR is already used, extensively,
to gather statistics about client-side values
such as user-provided URL domains
in the Chrome Web browser.

RAPPOR
can be easily applied to collect execution trace data,
while preserving both the privacy and anonymity of users.
In particular,
for reports about the frequency of system calls used by empirical programs
it is sufficient to utilize ``basic RAPPOR''---a simple, binary form of randomized response.
Such reports would be low overhead, 
both in terms of computation and their size, 
and can be easily aggregated into data-driven security policies.

\subsection{Matching User Expectations and Software Permissions}
At Google,
we have in the last few years
developed techniques for estimating
how users expect software to behave,
in different aspects.
This effort has been performed in the context of online software markets,
such as those that exist for mobile phone and Web platforms.
The goal has been to improve the
use of security-related permissions by software in thosse markets
by creating ``peer groups'' of similar software~\cite{applesAndOranges}.

Our premise has been that users will expect 
similar behavior
from software that they perceive to offer similar functionality.

Recently,
in work led by Martin Pelikan,
we have gotten very good results 
finding apparently-similar software
using modern machine-learning techniques---in
particular, 
by using \texttt{word2vec} skip-gram models
on the data available about different software,
such as its descriptions and how users' interact with the software in online markets.
As a result,
we have been able to craft quite accurate software peer groups
of software that provides similar functionality,
and thereby have established
a good basis of comparison 
from which to estimate users' expectation.

For the purposes of security,
we have found that
knowing how users expect software to behave
(or a good estimate thereof)
can provide many benefits.
In particular,
it allows actual, concrete software behavior---such as
that which might be provided by an empirical program---to
be compared against users' expectation.
If there is a (large) discrepancy,
a number of remedial approaches can be taken.
Most notably, the software developers can be notified,
and asked to remedy the situation;
other alternatives
include 
further automated or manual review of the software,
different handling of the software in the market,
or even asking users their opinion.
Some of those remedial approaches 
have been found suitable for practical application,
when measurements have proven them to have clear benefits to users.

\section{Conclusion}
When deciding whether software should be permitted to perform a security-relevant action,
it seems like a good idea to consider the historical evidence
of what actions that software has performed in the past.
For popular, widely-used software,
there are literally billions of executions from which to draw such historical evidence,
thereby allowing a very accurate view of what constitutes ``normal'' software execution to be established.
Furthermore,
for the most part,
this same software is already networked, 
and could provide execution trace data to online services that aggregated such evidence.

Motivated by the above,
a distinct model for data-driven software security can be established.
This data-driven model is different from the traditional approaches of anomaly and intrusion detection,
e.g., in its comprehensiveness and integration with software-development processes.
This model 
immediately raises concerns about efficiency, privacy, and practical utility,
but these concerns
can be positively addressed,
using existing techniques and mechanisms.

While many questions remain about the model's general applicability and deployment,
its enforcement could already provide substantial security benefits
to software that runs in some important domains---by reducing its attack surface,
and
thereby protecting the
software in the same manner as firewalls 
have very successfully protected networks.

\section*{Acknowledgment}
The ideas in this position paper have benefited from the thoughts and work of many.
In alphabetical order, thanks go to 
Mart{\'\i}n Abadi,
Blake Bassett,
Ludovico Cavedon,
Andy Chu, 
Giulia Fanti, 
Iulia Ion,
Suman Jana,
Noah Johnson, 
Aleksandra Korolova,
Karen Lees, 
Ilya Mironov, 
Martin Pelikan,
Vasyl Pihur, 
Ananth Raghunathan,
Sooel Son, 
Dawn Song,
Michael Vrable,
Moti Yung,
and
Yinqian Zhang.
Thanks also to Alessandra Gorla, Florian Gross, and Andreas Zeller
for collaboration based on their related work in~\cite{Chabada}
and discussion of their most recent work on mining sandboxes~\cite{MiningSandboxes}.

\bibliographystyle{IEEEtran}
\bibliography{IEEEfull,main}

\end{document}